\documentclass[aps,prb,twocolumn,groupedaddress,showpacs]{revtex4-1}

\usepackage{graphicx}
\usepackage{dcolumn}
\usepackage{bm}
\usepackage{hyperref}
\usepackage[mathlines]{lineno}

\usepackage{amsmath}

\begin{document}

\title{Spin Wave Excitations in AFe$_{1.5}$Se$_2$ (A=K, Tl): Analytical Study}
\author{Miao Gao$^{1}$}
\author{Xun-Wang Yan$^{1,2}$}
\author{Zhong-Yi Lu$^{1}$}\email{zlu@ruc.edu.cn}

\date{\today}

\affiliation{$^{1}$Department of Physics, Renmin University of
China, Beijing 100872, China}

\affiliation{$^{2}$School of Physics and Electrical Engineering, Anyang Normal University,
Anyang 455002, China}

\begin{abstract}

We have analytically solved the spin wave excitations for the intercalated ternary iron-selenide AFe$_{1.5}$Se$_2$ (A=K, Tl) in the $4\times 2$ collinear antiferromagnetic order. It is found that there are one acoustic branch (gapless Goldstone mode) and two gapful optical branches of spin wave excitations with each in double degeneracy. By examining the non-imaginary excitation frequency condition, we can determine the corresponding phase boundary. The exchange couplings between Fe moments in AFe$_{1.5}$Se$_2$ are derived based on the first-principles total energy calculations. The Fe spin is found to be $S=\frac{3}{2}$ through computing the antiferromagnetic quantum fluctuation. And it is further found that a very small spin-orientation anisotropy can remarkably suppress the antiferromagnetic quantum fluctuation. The spin dynamical structure factors are calculated and discussed in associated with neutron inelastic scattering experiment.

\end{abstract}


\maketitle

\section{Introduction}

The discovery of iron-based superconductors \cite{kamihara} has stimulated great interest on the investigation of unconventional superconducting mechanism, in which magnetism is considered to play a substantial role. It is well-known that the parent compound of a cuprate superconductor is an antiferromagnetic (AFM) Mott insulator with a checkerboard AFM order on the copper square lattice. In contrast, the parent compound of an iron-based superconductor was found to be an AFM semi-metal \cite{lu0} with either collinear \cite{cruz,ma1} or bi-collinear \cite{ma,bao,shi} AFM order on the iron square lattice. Regarding the nature of the magnetism, there are basically two contradictive views. The one \cite{mazin} is based on itinerant electron picture, in which the Fermi surface nesting is responsible for the collinear AFM order. On the contrary, the other one is based on local moment interactions which can be described by the $J_1$-$J_2$ frustrated Heisenberg model. \cite{yildirim,si,ma1} And it was further shown \cite{ma1} that the underlying driving force herein is the As-bridged AFM superexchange interaction between a pair of next-nearest-neighboring fluctuating Fe local moments embedded in itinerant electrons. There are now more and more evidences in favor of the fluctuating Fe local moment picture. Especially, the neutron inelastic scattering experiments have shown that the low-energy magnetic excitations can be well described by the spin waves based on the $J_1$-$J_2$ Heisenberg model.\cite{dai-0,dai-1,dai-2,dai-3}

The newly discovered intercalated ternary iron-selenide superconductors A$_y$Fe$_x$Se$_2$ (A=K, Tl)\cite{chen,Cs,fang} show rich phase diagrams and many unusual physical properties that have not been found in other iron-based superconductors, for example, the superconductivity was found to coexist with a strong AFM order with a giant magnetic moment of 3.31 $\mu_B$/Fe formed below a Neel temperature of 559$K$ \cite{muSR,bao1} and to be proximity to an AFM insulating phase.\cite{fang} These reveal the close relationship between unconventional superconductivity and antiferromagnetism once more, and have triggered another surge of interest for the investigation of unconventional superconducting mechanism.

The compounds A$_y$Fe$_x$Se$_2$ have the ThCr$_{2}$Si$_{2}$ type crystal structure, isostructural with 122-type iron pnictides.\cite{rotter} However, the stable structures of A$_y$Fe$_x$Se$_2$ contain Fe vacancies ordered in either $\sqrt{5}\times \sqrt{5}$ or $4\times 2$ superstructure due to the balance required in chemical valences, which respectively correspond to A$_{0.8}$Fe$_{1.6}$Se$_2$ with one-fifth Fe vacancies or AFe$_{1.5}$Se$_2$ with one-quarter Fe vacancies. For A$_{0.8}$Fe$_{1.6}$Se$_2$, the neutron observation has found that it has a $\sqrt{5}\times \sqrt{5}$ blocked checkerboard AFM order to match the Fe vacancy superstructure.\cite{bao1} For AFe$_{1.5}$Se$_2$, the first-principles electronic structure calculations predicted\cite{yan} that its ground state is in a $4\times 2$ collinear AFM order, as shown in Fig. \ref{fig:Mag}(d). Moreover, the calculations further showed \cite{yan,yan11} that both compounds AFe$_{1.5}$Se$_2$ and A$_{0.8}$Fe$_{1.6}$Se$_2$ are antiferromagnetic semiconductors with band gaps of dozens and hundreds meV, respectively. Such band gaps have been confirmed by the recent ARPES and transport measurements.\cite{arpes} It was further proposed that the parent compound of an A$_y$Fe$_x$Se$_2$ superconductor is an AFM semiconductor either A$_{0.8}$Fe$_{1.6}$Se$_2$ or AFe$_{1.5}$Se$_2$.\cite{yan,yan11} The latest neutron diffraction experiment shows that the parent compound is likely AFe$_{1.5}$Se$_2$ with a $4\times 2$ collinear AFM order.\cite{zhao} In order to well understand the magnetism in the compound AFe$_{1.5}$Se$_2$, we have studied the magnetic excitations and spin dynamical structure factors, which can be directly detected by neutron experiments. Our approach is based on the linearized spin wave theory upon the effective spin Heisenberg model.

\section{Effective model}

\begin{figure}[h]
\begin{center}
\includegraphics[width=8.0cm]{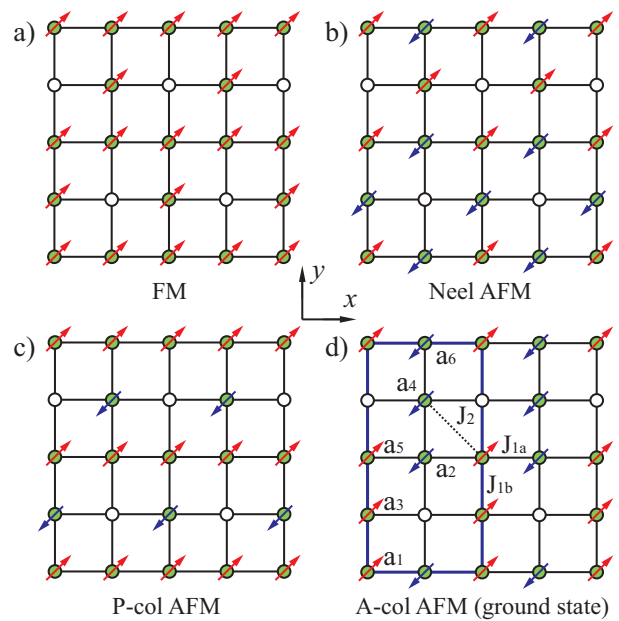}
\caption{(Color online) Different magnetic orders on an Fe-Fe square layer with one-quarter Fe vacancies ordered in rhombus: (a) ferromagnetic order; (b) N\'{e}el antiferromagnetic (AFM) order, in which all pairs of the nearest-neighboring Fe moments are in antiparallel order; (c) P-collinear AFM order, in which the Fe moments are in antiparallel order along the lines with the Fe vacancies; (d) A-collinear AFM order (ground state), in which the Fe moments are in antiparallel order along the lines without the Fe vacancies. The filled circles represent the Fe atoms while the empty circles represent the Fe vacancies. The red (blue) arrows represent the up-spins (down-spins). The A-collinear AFM order can be divided into six ferromagnetic sublattices, labeled successively as $a_1$ to $a_6$. $J_{1a}$ and $J_{1b}$ are the exchange couplings along the nearest neighbor Fe-Fe directions without and with the Fe vacancies, respectively. $J_2$ is the next-nearest neighbor exchange coupling. The rectangle enclosed by the thick blue solid lines denote a $4\times 2$ magnetic unit cell. The $x$ and $y$ axes are also shown.}
\label{fig:Mag}
\end{center}
\end{figure}

As schematically shown in Fig. \ref{fig:Mag}(d), in the ground state of the compound AFe$_{1.5}$Se$_2$,\cite{yan} the Fe vacancies are ordered in a rhombus structure, in which unit cell there are two inequivalent Fe atoms according to the number of neighboring Fe atoms, namely 2-Fe-neighbored and 3-Fe-neighbored Fe atoms respectively. The corresponding magnetic order is a $4\times 2$ collinear AFM order (also called A-col AFM order here), as shown in Fig. \ref{fig:Mag}(d), in which the Fe moments are antiferromagnetically ordered along the lines without Fe vacancies and ferromagnetically ordered along the lines perpendicular.

To describe the magnetism in the compound AFe$_{1.5}$Se$_2$, considering the Se-bridging effect,\cite{yan,ma1} we adopt the spin Heisenberg model with the nearest and next-nearest neighbor exchange couplings between the Fe moments with quantum spin $\vec{S}$ as follows,
\begin{equation}\label{eq:Ham}
\hat{H}=J_{1a}\sum_{\langle i,\delta_a \rangle}\vec{S}_i\cdot\vec{S}_{\delta_a} +J_{1b}\sum_{\langle i,\delta_b \rangle}\vec{S}_i\cdot\vec{S}_{\delta_b} + J_2\sum_{ \ll ij \gg}\vec{S}_i\cdot\vec{S}_j,
\end{equation}
whereas $\langle i,\delta_a \rangle$, $\langle i,\delta_b \rangle$, and $\ll ij \gg$ denote the summation over the nearest neighbors along the lines without Fe vacancies, the nearest neighbors along the lines with Fe vacancies, and the next-nearest neighbors, respectively. To be more specific, the exchange coupling $J_{1a}$ links a pair of 3-Fe-neighbored Fe atoms and the exchange coupling $J_{1b}$ links a 2-Fe-neighbored Fe atom and a 3-Fe-neighbored Fe atom, while the exchange coupling $J_2$ connects a 2-Fe-neighbored Fe atom with a 3-Fe-neighbored Fe atom (see Fig. \ref{fig:Mag}(d)).

\begin{figure}[h]
\begin{center}
\includegraphics[width=8.0cm]{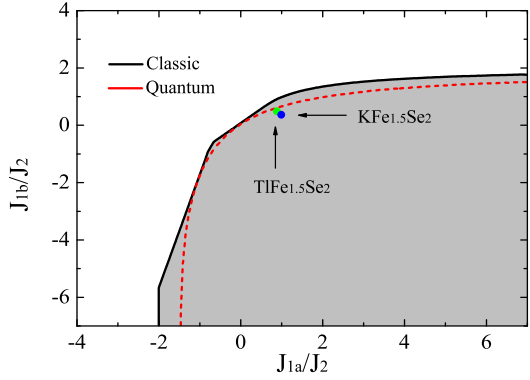}
\caption{(Color online) Phase diagram of A-collinear antiferromagnetic order, marked by the shadowed region, in variance with the exchange couplings $J_{1a}$, $J_{1b}$, and $J_2$, as defined in Fig. \ref{fig:Mag}(d). Black solid and red dashed lines represent the classical Monte Carlo calculation (Ref. \onlinecite{QMS}) and our result determined by examining the non-imaginary excitation frequency condition, respectively. Blue and green solid dots stand for the positions of KFe$_{1.5}$Se$_2$ and TlFe$_{1.5}$Se$_2$ respectively.}
\label{fig:diagram}
\end{center}
\end{figure}

The magnetic phase diagram of the Hamiltonian \eqref{eq:Ham} has been studied by classical Monte Carlo simulations in Ref. \onlinecite{QMS}. In the shadowed part as shown in Fig. \ref{fig:diagram}, the ground state of the Hamiltonian \eqref{eq:Ham} is in the A-col AFM phase with a $4\times 2$ collinear AFM order, whose spin dynamics will be studied below.

\section{Spin dynamics}

For a spin Heisenberg model with a long-range magnetic order in its ground state, the linearized spin wave theory with the Holstein-Primakoff (HP) transformation\cite{HP} is a standard approach to obtain the spin wave excitations and other dynamical properties. In the case of a simple magnetic unit cell, namely no more than two spins per cell, the approach can easily and directly give an analytical solution, for example, the well-known quadratic and linear dispersion behaviors in low energy for ferromagnetic and antiferromagnetic orders, respectively. However, for a complex magnetic structure with a magnetic unit cell containing more than two spins, it is still a severely challenging task to analytically solve the spin wave excitations even though lots of efforts have been devoted in the past.\cite{Meyer,Wallace,Zhang Zhi-dong,Milica} One thus has to be satisfied with numerical solutions. The underlying difficulty is as follows. After the linearized HP transformation, the spin Heisenberg model is transformed into a quadratic Bosonic Hamiltonian. Conventionally one attempts to construct a Bogoliubov transformation to diagonalize the Bosonic Hamiltonian.\cite{Bogoliubov} In general, there is no practical procedure to analytically construct such a Bogoliubov transformation for a Bosonic Hamiltonian with more than two component bosons, corresponding to a case of more than two spins per unit cell.

Nevertheless an analytical solution is usually in desire and very helpful to understand physics, especially more helpful in the case of a spin Heisenberg model in comparison with neutron measurement, for example, to determine the exchange couplings. Here by studying the Hamiltonian \eqref{eq:Ham}, rather than constructing a Bogoliubov transformation as conventionally, we are going to illustrate an alternative method, which is able to analytically solve spin wave excitations for the cases of multi-spin unit cells. This method employs the equation of motion to construct a secular equation, i.e. an algebraic equation. According to the algebraic basic theorem, an algebraic equation can be analytically solved up to the fourth power, which corresponds to an antiferromagnetic order with eight spins per unit cell. To our knowledge, the advantage of this method applied on a complex magnetic structure has not been well recognized.

\subsection{Spin wave excitations}

We now consider the Hamiltonian \eqref{eq:Ham} in the A-col AFM phase with a $4\times 2$ collinear AFM order. As shown in Fig. \ref{fig:Mag}(d), the magnetic unit cell contains two vacancies and six spins, i.e. three up spins and three down spins, which are labeled by $a_\xi$ with $\xi$ ($\xi=1,\ldots,6$) being odd number or even number, corresponding to up or down spins respectively. In other words, the original antiferromagnetic lattice can be divided into two vacancy-sublattices and six ferromagnetic sublattices, in each of which we can perform the linearized Holstein-Primakoff transformation for the quantum spin $\vec{S}$ at each site $i$ as follows,
\begin{equation}
\nonumber
\begin{array}{cc}
\left\{
\begin{array}{l}
S_i^+=\hat{a}_{\xi i}\sqrt{2S} \\
S_i^-=\hat{a}_{\xi i}^{\dagger}\sqrt{2S} \\
S_i^z=S-\hat{a}_{\xi i}^{\dagger}\hat{a}_{\xi i} \\
\xi\in \text{odd number}
\end{array}
\right.
&
~~~~
\left\{
\begin{array}{l}
S_i^+=\hat{a}_{\xi i}^{\dagger}\sqrt{2S} \\
S_i^-=\hat{a}_{\xi i}\sqrt{2S} \\
S_i^z=\hat{a}_{\xi i}^{\dagger}\hat{a}_{\xi i}-S \\
\xi\in \text{even number}
\end{array}
\right.
\end{array}
\end{equation}
where $\hat{a}_{\xi i}^{\dagger}$ ($\hat{a}_{\xi i}$) is the $\xi$-th component boson creation (annihilation) operator and $i$ is the site index in the $\xi$-th sublattice. We then perform the following Fourier transformation,
\begin{equation}
\nonumber
\begin{array}{cc}
\left\{
\begin{array}{l}
\hat{a}_{\xi i}=\frac{1}{\sqrt{N}}\sum_{\textbf{k}} e^{i\textbf{k}\cdot\textbf{R}_i}\hat{a}_{\xi \textbf{k}} \\
\hat{a}_{\xi i}^{\dagger}=\frac{1}{\sqrt{N}}\sum_{\textbf{k}} e^{-i\textbf{k}\cdot\textbf{R}_i}\hat{a}_{\xi\textbf{k}}^{\dagger} \\
\xi\in \text{odd number}
\end{array}
\right.
&
\left\{
\begin{array}{l}
\hat{a}_{\xi i}=\frac{1}{\sqrt{N}}\sum_\textbf{k} e^{-i\textbf{k}\cdot \textbf{R}_i}\hat{a}_{\xi \textbf{k}} \\
\hat{a}_{\xi i}^{\dagger}=\frac{1}{\sqrt{N}}\sum_\textbf{k} e^{i\textbf{k}\cdot \textbf{R}_i}\hat{a}_{\xi \textbf{k}}^{\dagger} \\
\xi\in \text{even number}
\end{array}
\right.
\end{array}
\end{equation}
where $\textbf{R}_i$ denotes the position vector of site $i$ in the $\xi$-th sublattice, $\textbf{k}=(k_x,k_y)$ is a wave vector in the magnetic Brillouin zone, and $N$ is the total site number of each sublattice.

It turns out that the spin Hamiltonian \eqref{eq:Ham} is now transformed into a quadratic Bosonic Hamiltonian in the momentum space as follows,\cite{explain}
\begin{equation}\label{eq:Fe1.5-H}
\begin{array}{l}
\hat{H}=E_0+\hat{H}_1 ,\\
E_0=(4J_{1b}-4J_{1a}-8J_2)NS(S+1) ,\\
\hat{H}_1=S\sum_\textbf{k}\hat{\psi}_\textbf{k}^\dag H_1^\textbf{k}\hat{\psi}_\textbf{k} , \\
\end{array}
\end{equation}
whereas
$$
\begin{array}{l}
H_1^\textbf{k}=
\left(
\begin{array}{cccccc}
D_1 & 0 & A_\textbf{k} & B_\textbf{k} & 0 & C_\textbf{k} \\
0 &  D_1 & B_\textbf{k} & A_\textbf{k} & C_\textbf{k} & 0 \\
A_\textbf{k}^* & B_\textbf{k}^* & D_2 & 0 & A_\textbf{k} & B_\textbf{k} \\
B_\textbf{k}^* & A_\textbf{k}^* & 0 & D_2 & B_\textbf{k} & A_\textbf{k} \\
0 & C_\textbf{k} & A_\textbf{k}^* & B_\textbf{k}^* & D_1 & 0 \\
C_\textbf{k} & 0 & B_\textbf{k}^* & A_\textbf{k}^* & 0 & D_1 \\
\end{array}
\right), \\
\begin{array}{l}
D_1=-J_{1b}+2J_{1a}+2J_2 ,\\
D_2=-2J_{1b}+4J_2 ,\\
A_\textbf{k}=J_{1b}e^{i\theta_y}, \\
B_\textbf{k}=2J_2\cos(\theta_x)e^{-i\theta_y} , \\
C_\textbf{k}=2J_{1a}\cos(\theta_x) ,\\
\theta_x=k_xl , ~~ \theta_y=k_yl ,\\
\hat{\psi}_\textbf{k}^\dag=(\hat{a}_{1\textbf{k}}^{\dag}~\hat{a}_{2\textbf{k}}~
\hat{a}_{3\textbf{k}}^{\dag}
~\hat{a}_{4\textbf{k}}~\hat{a}_{5\textbf{k}}^{\dag}~\hat{a}_{6\textbf{k}}) , \\
\end{array}
\end{array}
$$
where $l$ is the length of two nearest neighbor sites, namely the Fe-Fe bond length. In the following we take $l$ as the length unit for convenience.

Next we construct a secular equation to solve the eigenvalues of the Hamiltonian \eqref{eq:Fe1.5-H} by using the equation of motion method. We first consider the following linear combination,
$$
\hat{\alpha}_{\textbf{k}}=-\sum_{\xi\in \text{odd}} u_{\textbf{k}\xi}\hat{a}_{\xi \textbf{k}}^\dag+
\sum_{\xi\in \text{even}} u_{\textbf{k}\xi}\hat{a}_{\xi \textbf{k}} ,
$$
or
$$
\hat{\beta}_{\textbf{k}}=\sum_{\xi\in \text{odd}} v_{\textbf{k}\xi}\hat{a}_{\xi \textbf{k}}-
\sum_{\xi\in \text{even}} v_{\textbf{k}\xi}\hat{a}_{\xi \textbf{k}}^\dag  .\\
$$
Without loss of generality, the coefficients $u_{\textbf{k}\xi}$ and $v_{\textbf{k}\xi}$ are assumed to be complex numbers. Here $\hat{\alpha}_{\textbf{k}}$ and $\hat{\beta}_{\textbf{k}}$ are the boson operators, which satisfy the relationship of the boson commutation, namely $[\hat{\alpha}_{\textbf{k}}, \hat{\alpha}_{\textbf{k}'}^\dag]=\delta_{\textbf{k}\textbf{k}'}$ and $[\hat{\beta}_{\textbf{k}}, \hat{\beta}_{\textbf{k}'}^\dag]=\delta_{\textbf{k}\textbf{k}'}$. This imposes the following constraints upon the coefficients $u_{\textbf{k}\xi}$ and $v_{\textbf{k}\xi}$,
$$
\begin{array}{r}
-\sum_{\xi\in \text{odd}} u_{\textbf{k}\xi}u_{\textbf{k}\xi}^*+
\sum_{\xi\in \text{even}} u_{\textbf{k}\xi}u_{\textbf{k}\xi}^*=1 , \\
\sum_{\xi\in \text{odd}} v_{\textbf{k}\xi}v_{\textbf{k}\xi}^*-
\sum_{\xi\in \text{even}} v_{\textbf{k}\xi}v_{k\xi}^*=1  ,\\
\end{array}
$$
whereas $*$ means the complex conjugate and remains the same meaning as follows.

Now we assume that $\hat{\alpha}_{\textbf{k}}$ or $\hat{\beta}_{\textbf{k}}$ describes an eigen-mode of the Hamiltonian \eqref{eq:Fe1.5-H}, it is then required that $\hat{\alpha}_{\textbf{k}}$ or $\hat{\beta}_{\textbf{k}}$ fulfils the
following equations of motion respectively,
$$
i\hbar\dot{\hat{\alpha}}_{\textbf{k}}=[\hat{\alpha}_{\textbf{k}}, \hat{H}]=\lambda_\textbf{k} \alpha_{\textbf{k}} \quad
\text{or} ~~~
i\hbar\dot{\hat{\beta}}_{\textbf{k}}=[\hat{\beta}_{\textbf{k}}, \hat{H}]=\lambda_\textbf{k}' \hat{\beta}_{\textbf{k}} ,
$$
where $\lambda_\textbf{k}$ and $\lambda_\textbf{k}'$ are the respective eigenvalues.
This gives rise to a generalized eigenvalue problem at each wave vector $\textbf{k}$ in the magnetic Brillouin zone,
\begin{equation}
\label{eq:alpha-eq}
H_1^\textbf{k} U_\textbf{k}=-\lambda_\textbf{k}\sigma_3U_\textbf{k}, \quad  U_\textbf{k}=(u_{\textbf{k}1}, u_{\textbf{k}2}, u_{\textbf{k}3}, u_{\textbf{k}4}, u_{\textbf{k}5}, u_{\textbf{k}6})^T ,\\
\end{equation}
or
\begin{equation}
\label{eq:beta-eq}
H_1^{\textbf{k}*} V_\textbf{k}=\lambda_\textbf{k}'\sigma_3V_\textbf{k},  \quad V_\textbf{k}=(v_{\textbf{k}1}, v_{\textbf{k}2}, v_{\textbf{k}3}, v_{\textbf{k}4}, v_{\textbf{k}5}, v_{\textbf{k}6})^T ,\\
\end{equation}
where the $6\times 6$ matrix $H_1^{\textbf{k}}$ is given in Eq. \eqref{eq:Fe1.5-H}, and $\sigma_3$ is a $6\times 6$ diagonal matrix with diagonal elements being (1, -1, 1, -1, 1, -1) respectively.

Eqs. \eqref{eq:alpha-eq} and \eqref{eq:beta-eq} will show that each eigenvalue of the Hamiltonian \eqref{eq:Fe1.5-H} is in double degeneracy, which reflects the symmetry or equivalence between the spin-up and spin-down in the ground state. For convenience, we only deal with  Eq. \eqref{eq:alpha-eq} in detail. We will not diagonalize Eq. \eqref{eq:alpha-eq} directly since this yields a set of quadratic equations with at least 36 unknowns, which is unlikely to give an analytical expression. Instead, we come to solve the secular equation of the Hamiltonian \eqref{eq:alpha-eq}, namely zeroing the determinant of the matrix $H_1^{\bf{k}}+\lambda_\textbf{k}\sigma_3$, which is the condition for there being a nonzero vector $U_\textbf{k}$. Actually, the Hamiltonian \eqref{eq:beta-eq} has the same secular equation. The secular equation turns out to be an algebraic equation with one unknown $\lambda_\textbf{k}$, namely
\begin{equation}\label{eq:cubic}
\lambda_\textbf{k}^6+b\lambda_\textbf{k}^4+c\lambda_\textbf{k}^2+d=0,
\end{equation}
whereas the coefficients $b$, $c$, and $d$ all are real functions of the wave vector $\textbf{k}$, and composed of the exchange couplings. For the detailed expressions on $b$, $c$, and $d$, please refer to Appendix I.

Setting $\gamma=\lambda_\textbf{k}^2$, we then have $\gamma^3+b\gamma^2+c\gamma+d=0$ from Eq. \eqref{eq:cubic}. Further setting $\gamma=\alpha-\frac{b}{3}$, the equation reduces to $\alpha^3+p\alpha+q=0$,
whereas $p=-\frac{b^2}{3}+c$, and $q=\frac{2b^3}{27}-\frac{bc}{3}+d$.
When the discriminant $\Delta=[(\frac{q}{2})^2+(\frac{p}{3})^3] \leq 0$, there are three real roots for this cubic equation, namely
$
\alpha_1=2\sqrt[3]{r}\cos\phi ,~
\alpha_2=2\sqrt[3]{r}\cos(\phi+\frac{2\pi}{3}) ,~
\alpha_3=2\sqrt[3]{r}\cos(\phi+\frac{4\pi}{3}) ,
$
whereas $r=\sqrt{-(\frac{p}{3})^3}$ and $\phi=\frac{1}{3}\arccos(-\frac{q}{2r})$. Now we arrive at the condition, which makes all six roots of Eq. \eqref{eq:cubic} being real, is that $\alpha_1 -\frac{b}{3} \ge 0$ and $\alpha_2-\frac{b}{3}\ge 0$ and $\alpha_3-\frac{b}{3}\ge 0$.
In the end, we analytically obtain six branches of eigenvalues for the Hamiltonian \eqref{eq:Fe1.5-H}, in three pairs of positive and negative, similar to the case of calculating lattice phonon excitations. And we keep the three positive branches to describe the spin wave excitations as follows,
\begin{equation}\label{spinwave}
\left\{
\begin{array}{l}
\lambda_{1\textbf{k}}=\hbar\omega_1(\textbf{k})=S\sqrt{\alpha_1-b/3} ,\\
\lambda_{2\textbf{k}}=\hbar\omega_2(\textbf{k})=S\sqrt{\alpha_2-b/3} ,\\
\lambda_{3\textbf{k}}=\hbar\omega_3(\textbf{k})=S\sqrt{\alpha_3-b/3} .\\
\end{array}
\right.
\end{equation}

Likewise we can analytically solve Eq. \eqref{eq:beta-eq} to obtain the same set of eigenvalues, but with the different eigenvectors. Lets take the complex conjugate of Eq. \eqref{eq:alpha-eq}, namely $H_1^{\textbf{k}*}U_\textbf{k}^*=-\lambda_\textbf{k}\sigma_3U_\textbf{k}^*$. Comparing with Eq. \eqref{eq:beta-eq}, we can see that a positive or negative eigenvalue ($\pm\lambda_\textbf{k}$) with eigenvector $U_\textbf{k}$ of Eq. \eqref{eq:alpha-eq} is a negative or positive eigenvalue ($\mp\lambda_\textbf{k}$) with eigenvector $V_{\textbf{k}}=U_\textbf{k}^*$ of Eq. \eqref{eq:beta-eq}. This directly shows that the spin wave excitations are in double degeneracy.

The Hamiltonian \eqref{eq:Fe1.5-H} is derived from the spin Hamiltonian \eqref{eq:Ham} on the assumption of the ground state being in the $4\times 2$ collinear AFM order (see Fig. \ref{fig:Mag}(d)). As shown above, the eigenvalues of the Hamiltonian \eqref{eq:Fe1.5-H} may be imaginary. Such a case, if happening, indicates that the $4\times 2$ collinear AFM order is unstable for the spin Hamiltonian \eqref{eq:Ham}. This is similar to the case of imaginary phonon vibration modes, which indicates that the corresponding crystal structure is unstable. Thus we can determine the boundary of A-col phase in the phase diagram of the spin Hamiltonian \eqref{eq:Ham} by checking the condition whether the eigenvalues of the Hamiltonian \eqref{eq:Fe1.5-H} being real or not. Figure \ref{fig:diagram} shows our calculated phase boundary by the red dashed line, which is quite close to the one given by the classical Monte Carlo simulations. The difference between the two is that the AFM quantum fluctuations are included in our spin wave calculations but not in the classical Monte Carlo simulations.

In order to calculate physical quantities, we need to know the eigenstates corresponding to the spin wave excitations, which but are not directly $U_\textbf{k}$ and $V_\textbf{k}$ of Eqs. \eqref{eq:alpha-eq} and \eqref{eq:beta-eq} because of the boson's characteristic. Let us denote the eigenvectors corresponding to the positive and negative eigenvalues of Eq. \eqref{eq:alpha-eq} by $U_{\eta,\textbf{k}}^{(+)}$ and $U_{\eta,\textbf{k}}^{(-)}$ respectively, with $\eta=1,2,3$ being the branch index of the spin wave excitations. Then the Hamiltonian \eqref{eq:Fe1.5-H} will be diagonalized into
\begin{equation}
\nonumber
H=E_0+\sum_{\eta \bf{k}}\hbar\omega_\eta(\hat{\alpha}_{\eta \bf{k}}^\dag\hat{\alpha}_{\eta \bf{k}}
+\hat{\beta}_{\eta \bf{k}}^\dag\hat{\beta}_{\eta \bf{k}}+1)
\end{equation}
after the following linear transformation,
\begin{equation}
\nonumber
\left\{
\begin{array}{l}
\hat{\alpha}_{\eta \bf{k}}=-\sum_{\xi\in \text{odd}} u_{\eta \xi,\textbf{k}}^{(+)} \hat{a}_{\xi \bf{k}}^\dag+
\sum_{\xi\in \text{even}} u_{\eta \xi,\textbf{k}}^{(+)} \hat{a}_{\xi \bf{k}}, \\
\hat{\beta}_{\eta \bf{k}}^\dag=\sum_{\xi\in \text{odd}} u_{\eta \xi,\textbf{k}}^{(-)} \hat{a}_{\xi \bf{k}}^\dag-
\sum_{\xi\in \text{even}} u_{\eta \xi,\textbf{k}}^{(-)} \hat{a}_{\xi \bf{k}}. \\
\end{array}
\right.
\end{equation}
The above set of equations can be rewritten in a matrix form as
\begin{equation}
\label{eq:eigenstates}
(\hat{\alpha}_{1\bf{k}}~\hat{\beta}_{1\bf{k}}^{\dag}~ \hat{\alpha}_{2\bf{k}}~\hat{\beta}_{2\bf{k}}^{\dag} ~ \hat{\alpha}_{3\bf{k}}~\hat{\beta}_{3\bf{k}}^{\dag})
=(\hat{a}_{1\bf{k}}^{\dag}~\hat{a}_{2\bf{k}}~ \hat{a}_{3\bf{k}}^{\dag}~\hat{a}_{4\bf{k}} ~ \hat{a}_{5\bf{k}}^{\dag}~\hat{a}_{6\bf{k}})Q_{\textbf{k}}.
\end{equation}
Here the six columns of the matrix $Q_{\textbf{k}}$ are nothing but the eigenstates corresponding to the spin wave excitations.

\subsection{Spin dynamical structure factor}

As shown above, $S^z$ just contributes to elastic scattering rather
than inelastic scattering since it does not change the number of
magnons. In contrast, $S^x$ and $S^y$ contribute to inelastic
scattering by changing the number of magnons. In zero temperature
the spin dynamical structure factor (SDSF) in inelastic scattering
process through single magnon excitations is defined as
\begin{equation}
\label{eq:dynamic factor}
\begin{array}{lll}
S(\textbf{k},\omega)& = & \sum_f\sum_{i=x,y}|\langle f|S^i(\textbf{k})|0\rangle|^2\delta(\omega-\omega_f) \\
 & = & S\sum_{\gamma}|Q^{-1}_{\gamma,\textbf{k}}|^2\delta(\omega-\omega_f),
\end{array}
\end{equation}
where $|0\rangle$ is the vacuum state, $|f\rangle$ denotes the final
states of a spin system with excitation energy $\omega_f$, and
$Q^{-1}_{\gamma,\textbf{k}}$ means the sum of the elements in the $\gamma$-th row of
the matrix $Q^{-1}_{\textbf{k}}$ given in Eq. \eqref{eq:eigenstates}.

\subsection{Sublattice magnetization}

In the $\xi$-th sublattice the AFM quantum fluctuation reduces the staggered magnetization
from its classical value $S$ by the following quantity $\Delta S_\xi$,
\begin{equation}
\label{eq:submag}
\begin{array}{ll}
\Delta S_\xi=S-\frac{1}{N}\langle\sum_{i=1}^N(S-\hat{a}_{\xi i}^\dag \hat{a}_{\xi i})\rangle=
\frac{1}{N}\langle\sum_{\bf{k}} \hat{a}_{\xi\bf{k}}^\dag \hat{a}_{\xi\bf{k}}\rangle \\
=\left\{
\begin{array}{ll}
\frac{1}{N}\sum_{\bf{k}}\sum_{\nu=1,3,5}|Q^{-1}_{\nu\xi,\textbf{k}}|^2, \quad\text{for} ~ \xi \in \text{odd number}, \\
\frac{1}{N}\sum_{\bf{k}}\sum_{\nu=2,4,6} |Q^{-1}_{\nu\xi,\textbf{k}}|^2, \quad\text{for} ~ \xi \in \text{even number}. \\
\end{array}
\right.
\end{array}
\end{equation}
where $Q^{-1}_{\nu\xi,\textbf{k}}$ denotes the element of the matrix $Q^{-1}_{\textbf{k}}$ at the crossing of the $\nu$-th row and the $\xi$-th column. The summation of $\bf{k}$ can be replaced by a two dimensional integral over the whole Brillouin Zone. It is well-known that the AFM quantum fluctuation is strongly influenced by the spatial dimension and the number of nearest neighbors of a site, namely the coordination number. There are two kinds of two Fe atoms (sites) in a magnetic unit cell in AFe$_{1.5}$Se$_2$, one is 2-Fe-neighbored and the other is 3-Fe-neighbored. Thus we need to calculate the AFM quantum fluctuation on these two kinds of sites separately. Moreover, there usually exists small spin-orientation anisotropy in realistic materials. To account for such anisotropy, one standard approach is to add the term of $-\Lambda_{z}\sum S_i^{z2}$ into the Hamiltonian \eqref{eq:Ham}. This turns out to add $2\Lambda_z$ to all diagonal term of $H_1^{\bf{k}}$ and $-3\Lambda_zNS(S+1)$ to $E_0$ in Eq. \eqref{eq:Fe1.5-H} respectively.

\section{Results and discussion}

The above analytical study will be very helpful both experimentally and theoretically. Experimentally in comparison with neutron inelastic scattering on AFe$_{1.5}$Se$_2$, we can determine the exchange couplings $J_{1a}$, $J_{1b}$, and $J_2$ through the spin wave
excitations (Eqs. \eqref{spinwave} and \eqref{eq:eigenstates}). This will help us better understand the magnetism in AFe$_{1.5}$Se$_2$. On the other hand, these exchange couplings can be theoretically derived from the relative energies of different magnetic states with respect to the non-magnetic state. For the detailed derivation, please refer to the appendix in Ref. \onlinecite{ma1}. To be specific, we need to first obtain total energies of the four different magnetic states, i.e. the ferromagnetic state ($E_{F}$), N\'{e}el AFM state ($E_N$), P-collinear AFM state ($E_P$), and A-collinear AFM state ($E_A$), as shown respectively in Fig. \ref{fig:Mag}. The energy differences among these different states result in three linearly independent equations as follows,
\begin{equation}
\nonumber
\left\{
\begin{array}{ll}
E_{N}-E_{F}=-4(J_{1a}+J_{1b})/3 ,\\
E_{A}-E_{N}=4(J_{1b}-2J_2)/3 ,\\
E_{A}-E_{P}=-4(J_{1a}-J_{1b})/3 , \\
\end{array}
\right.
\end{equation}
from which the exchange couplings $J_{1a}$, $J_{1b}$, and $J_2$ can be uniquely determined.

To obtain the total energies, we have carried out the first-principles electronic structure calculations on the compounds AFe$_{1.5}$Se$_2$, which were reported in Ref. \onlinecite{yan}. The calculated total energies and subsequently derived exchange couplings for the compounds AFe$_{1.5}$Se$_2$ are listed in Table \ref{table:energy}. Accordingly, in Fig. \ref{fig:diagram} we mark the positions of AFe$_{1.5}$Se$_2$ in the magnetic phase diagram, both of which are in the A-col phase, but close to the phase boundary. Thus the ground states of AFe$_{1.5}$Se$_2$ both are in the $4\times 2$ collinear AFM order, as observed in the neutron elastic scattering experiment, but they may be readily destructed by doping or applying pressure.

\begin{table}
\caption{Calculated energies of the four different magnetic states, i.e. the ferromagnetic state ($E_{F}$), N\'{e}el antiferromagnetic state ($E_N$), P-collinear antiferromagnetic state ($E_P$), and A-collinear antiferromagnetic state ($E_A$), as shown in Fig. \ref{fig:Mag}, for AFe$_{1.5}$Se$_2$ (A=K, Tl) (unit: meV/Fe). The energy of the nonmagnetic state is set to zero. The calculated exchange couplings $J_{1a}$, $J_{1b}$, and $J_2$ are also listed (unit: meV/S$^2$).}
\label{table:energy}
\begin{tabular}{|c|c|c|c|c|c|c|c|}
  \hline
  Compounds & $E_F$ & $E_N$ & $E_P$ & $E_A$ & $J_{1a}$ & $J_{1b}$ & $J_2$\\
  \cline{1-8}
  KFe$_{1.5}$Se$_2$ & -157.1 & -253.0 & -325.3 & -370.0 & 52.7 & 19.2 & 53.5\\
  \cline{1-8}
  TlFe$_{1.5}$Se$_2$ & -79.4 & -176.0 & -255.7 & -283.8 & 46.8 & 25.7 & 53.3\\
  \hline
\end{tabular}
\end{table}

From Table \ref{table:energy}, we also see that the exchange couplings $J_2$ are almost the same for both KFe$_{1.5}$Se$_2$ and TlFe$_{1.5}$Se$_2$, but the former is more anisotropic than the latter on the exchange couplings $J_1$, namely the difference between $J_{1a}$ and $J_{1b}$. Such a magnetic anisotropy will help lessen the AFM quantum fluctuations. With the help of Eq. \eqref{eq:submag}, we are now able to calculate the AFM quantum fluctuations in AFe$_{1.5}$Se$_2$. We summarize the calculated results in Table \ref{table:quantum}. As we find, the amount of the AFM quantum fluctuation in KFe$_{1.5}$Se$_2$ is less than the one in TlFe$_{1.5}$Se$_2$ by about 20\%. From Table II, we further find that the AFM quantum fluctuation on a 2-Fe-neighbored Fe atom is, as expected, rather larger than the one on a 3-Fe-neighbored atom, by about 10\% and 20\% for KFe$_{1.5}$Se$_2$ and TlFe$_{1.5}$Se$_2$ respectively. Moreover a very small spin-orientation anisotropy can remarkably reduce the AFM quantum fluctuations, as shown in the case of $\Lambda_z=0.02J_2$, in which the AFM quantum fluctuation is already reduced from 0.2521 (0.3162) to 0.1943 (0.2337) for KFe$_{1.5}$Se$_2$ (TlFe$_{1.5}$Se$_2$) by approximately 20\%. Experimentally the neutron measurements have shown that there is a small spin-orientation anisotropy in KFe$_{1.5}$Se$_2$ and the other iron pnictides.\cite{cruz,bao,shi,bao1,zhao} In addition, the effective ordering moment in KFe$_{1.5}$Se$_2$ was found to be 2.8$\mu_B$ per Fe atom.\cite{zhao} Here setting $S=\frac{3}{2}$, our calculations show that the ordering moment is 2.61$\mu_B$ with the spin anisotropy of $\Lambda_z=0.02J_2$ for KFe$_{1.5}$Se$_2$, in good agreement with the neutron observation. We thus arrive with $S=\frac{3}{2}$ for each Fe atom in KFe$_{1.5}$Se$_2$. In the following discussion, we set $S=\frac{3}{2}$ for each lattice site.

\begin{table}
\caption{Antiferromagnetic quantum fluctuations for AFe$_{1.5}$Se$_2$ (A=K, Tl) with and without spin anisotropy respectively. $\Delta S_{2n}$ and $\Delta S_{3n}$ denote the antiferromagnetic quantum fluctuation on a 2-Fe-neighbored and 3-Fe-neighbored Fe atom respectively. There are two 2-Fe-neighbored and four 3-Fe-neighbored Fe atoms in a $4\times 2$ magnetic unit cell (see Fig. \ref{fig:Mag}(d)). $\Delta S$ denotes the average antiferromagnetic quantum fluctuation on an Fe atom.} \label{table:quantum}
\begin{tabular}{|c|c|c|c|c|c|c|}
  \hline
  Compounds & \multicolumn{3}{c|}{$\Lambda_z=0.0$}
  &\multicolumn{3}{c|}{$\Lambda_z=0.02J_2$} \\
  \cline{2-7}
     & $\Delta S_{2n}$ & $\Delta S_{3n}$ & $\Delta S$ & $\Delta S_{2n}$ & $\Delta S_{3n}$ & $\Delta S$ \\
  \cline{1-7}
  KFe$_{1.5}$Se$_2$ & 0.2685 & 0.2439 & 0.2521 & 0.2093 & 0.1868 & 0.1943 \\
  \cline{1-7}
  TlFe$_{1.5}$Se$_2$ & 0.3665 & 0.2911 & 0.3162 & 0.2780 & 0.2115 & 0.2337 \\
  \hline
\end{tabular}
\end{table}

\begin{figure}[h]
\begin{center}
\includegraphics[width=8.0cm]{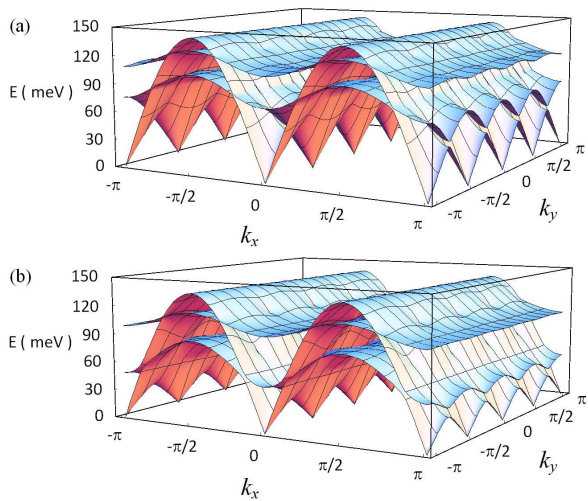}
\caption{(Color online) Three branches of the spin wave excitations in A-collinear antiferromagnetic phase with a $4\times 2$ magnetic unit cell (see Fig. \ref{fig:Mag}(d)). (a) KFe$_{1.5}$Se$_2$, in which $J_{1a}=23.4$, $J_{1b}=8.5$, and $J_2=23.8$ meV ($S=\frac{3}{2}$). (b) TlFe$_{1.5}$Se$_2$, in which $J_{1a}=20.8$, $J_{1b}=11.4$, and $J_2=23.7$ meV ($S=\frac{3}{2}$). Note $k_x$ is of spin-antiparallel direction without the Fe vacancies aligned and $k_y$ is of spin-parallel direction with the Fe vacancies aligned. Here the Fe-Fe bond length $l$ is taken as the length unit for convenience.}
\label{fig:spectrum}
\end{center}
\end{figure}

For an AFM structure with a magnetic unit cell containing six spins, physically there are six branches of spin wave excitations with every two in degeneracy, namely three 2-degenerated branches, among which one is the gapless Goldstone mode and the other two are gapful optical modes. With Table \ref{table:energy} and Eqs. \eqref{spinwave}, we can plot the spin wave spectra in the extended Brillouin zone for the compound AFe$_{1.5}$Se$_2$ in Fig. \ref{fig:spectrum}. As we see, there are exactly one gapless Goldstone mode and two gapful optical modes, and each branch of spin wave excitation is much more dispersive along the spin-antiparallel direction than along the spin-parallel direction. This is attributed to the Fe vacancies hindering the spin wave propagations, which are located along the spin-parallel direction rather than the spin-antiparallel direction (see Fig. \ref{fig:Mag}(d)). Furthermore, such hindering becomes more severe, also the spin wave diffraction becomes weaker, for the higher spin wave excitations because of the corresponding wavelengthes being shorter. Especially, the highest optical mode is almost dispersionless along the spin-parallel direction.

\begin{figure}[h]
\begin{center}
\includegraphics[width=8.0cm]{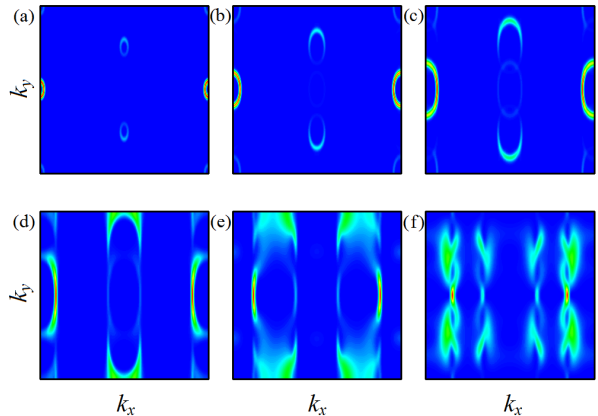}
\caption{(Color online) Constant-energy cuts of the spin dynamical structure factor (SDSF) for KFe$_{1.5}$Se$_2$, in which $J_{1a}=23.4$, $J_{1b}=8.5$, and $J_2=23.8$ meV ($S=\frac{3}{2}$). (a)-(f) are the SDSF at the excitation energies of 20, 40, 60, 80, 100, and 120 meV, respectively. $k_x$ and $k_y$ both vary from -$\pi$ to $\pi$. Note $k_x$ is of spin-antiparallel direction and $k_y$ is of spin-parallel direction.}
\label{fig:DSF-K}
\end{center}
\end{figure}

\begin{figure}[h]
\begin{center}
\includegraphics[width=8.0cm]{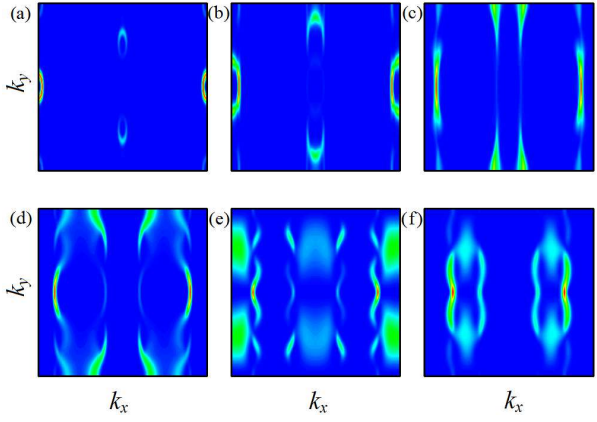}
\caption{(Color online) Constant-energy cuts of the spin dynamical structure factor (SDSF) for TlFe$_{1.5}$Se$_2$, in which $J_{1a}=20.8$, $J_{1b}=11.4$, and $J_2=23.7$ meV ($S=\frac{3}{2}$). (a)-(f) are the SDSF at the excitation energies of 20, 40, 60, 80, 100, and 120 meV, respectively. $k_x$ and $k_y$ both vary from -$\pi$ to $\pi$. Note $k_x$ is of spin-antiparallel direction and $k_y$ is of spin-parallel direction.}
\label{fig:DSF-Tl}
\end{center}
\end{figure}

To compute the spin dynamical structure factor by using Eq. \eqref{eq:dynamic factor}, the function $\delta(\omega-\omega_f)$ is smoothed by a Gaussian distribution function, which is centered at the excitation energy $\omega_f$ with the half-width at the half-maximum (HWHM) assigned to $0.2J_2$. We plot the spin dynamical structure factor at a constant cut energy, ranging from 20 to 120 meV for both compounds, as shown in Figs. 4 and 5 respectively.

For KFe$_{1.5}$Se$_2$, referring to Fig. 3, one can see that the diffraction peaks come from the acoustic (Goldstone mode) spin waves for the excitation energy $\omega_f<$ 50 meV, the
acoustic and the first optical spin waves for 50 meV $< \omega_f <$ 100 meV, and the first and second optical spin waves for $\omega_f >$ 100 meV, respectively. When $\omega_f=20$ meV, Figure \ref{fig:DSF-K} shows that the strongest diffraction peaks are located at ($\pm\pi$, 0). In addition there are two small half elliptical diffraction peaks near $\Gamma$ point and four almost invisible diffraction peaks at the four corners of the extended Brillouin zone. When $\omega_f$ increases from 20 to 80 meV, these ellipses extend larger. At 100 meV, the diffraction peaks become line-shaped. When $\omega_f=120$ meV, the diffraction peaks become helical lines along the $k_y$ directions, namely spin-parallel direction. These features are consistent with the strong anisotropy of the spin wave excitations along the spin-parallel and spin-antiparallel directions due to the Fe vacancies.

For TlFe$_{1.5}$Se$_2$, we also find the quite similar features on the spin dynamical structure factor (see Fig. \ref{fig:DSF-Tl}). Since the spin wave dispersions become further weaker along the spin-parallel direction in TlFe$_{1.5}$Se$_2$ than in KFe$_{1.5}$Se$_2$, as shown in Fig. 3, the diffraction peaks thus become line-shaped along $k_y$ at 60 meV rather than 100 meV for TlFe$_{1.5}$Se$_2$. Moreover, when $\omega_f$ increases from 80 to 120 meV, the diffraction peaks become pear-shaped patterns. The large green areas near the right and left boundaries of the extended Brillouin zone at 100 meV come from the flatness of the highest optical branch.

\section{Conclusion}

We have shown that we can analytically solve the spin wave excitations for a complex magnetic structure by using the equation of motion method in the framework of the linearized spin wave theory, which is illustrated by studying the magnetism of the newly discovered intercalated ternary iron-selenide AFe$_{1.5}$Se$_2$ (A=K, Tl). We find that there are one acoustic branch (gapless Goldstone mode) and two gapful optical branches of the spin wave excitations with each in double degeneracy in AFe$_{1.5}$Se$_2$. The phase boundary of AFe$_{1.5}$Se$_2$ in the $4\times 2$ collinear antiferromagnetic order is determined by examining the non-imaginary excitation frequency condition, which incorporates the antiferromagnetic quantum fluctuations. We also derive the exchange couplings between the Fe moments based on the first-principles total energy calculations, so that we can calculate and discuss the spin dynamical structure factors in connection with neutron inelastic scattering experiment. By computing the antiferromagnetic quantum fluctuations, we find that the Fe spin is $S=\frac{3}{2}$ in AFe$_{1.5}$Se$_2$ and a very small spin-orientation anisotropy can remarkably suppress the antiferomagnetic quantum fluctuations.

\begin{acknowledgements}

This work is supported by National Program for Basic Research of MOST of China (Grant No. 2011CBA00112) and National Natural Science Foundation of China (Grant Nos. 11190024 and 91121008).

\end{acknowledgements}

\section{Appendix I: coefficients in Eq. (5)}

\begin{widetext}
$$
\left\{
\begin{array}{l}
b=-4J_{1b}^2+16J_2^2\cos(\theta_x)^2+2C_{\bf{k}}^2-2D_1^2-D_2^2 ,\\
\\
c=4J_{1b}^4-32J_{1b}^2J_2^2\cos(\theta_x)^2+64J_2^4\cos(\theta_x)^4-4J_{1b}^2C_{\bf{k}}^2
+16J_2^2\cos(\theta_x)^2C_{\bf{k}}^2+C_{\bf{k}}^4+4J_{1b}^2D_1^2-16J_2^2\cos(\theta_x)^2D_1^2 \\
\qquad -2C_{\bf{k}}^2D_1^2+D_1^4+16J_{1b}J_2\cos(\theta_x)C_{\bf{k}}D_2-4J_{1b}^2D_1D_2
-16J_2^2\cos(\theta_x)^2D_1D_2-2C_{\bf{k}}^2D_2^2+2D_1^2D_2^2 ,\\
\\
d=2J_{1b}^4\cos(4\theta_y)C_{\bf{k}}^2+2J_{1b}^4C_{\bf{k}}^2+
16J_{1b}^2J_2^2\cos(\theta_x)^2\cos(4\theta_y)C_{\bf{k}}^2
-16J_{1b}^2J_2^2\cos(\theta_x)^2C_{\bf{k}}^2+32J_2^4\cos(\theta_x)^4\cos(4\theta_y)C_{\bf{k}}^2 \\ \qquad
-32J_{1b}^2J_2^2\cos(\theta_x)^2C_{\bf{k}}^2+32J_2^4\cos(\theta_x)^4C_{\bf{k}}^2
-16J_{1b}^3J_2\cos(\theta_x)\cos(4\theta_y)C_{\bf{k}}D_1+16J_{1b}^3J_2\cos(\theta_x)C_{\bf{k}}D_1 \\ \qquad
-64J_{1b}J_2^3\cos(\theta_x)^3\cos(4\theta_y)C_{\bf{k}}D_1
+64J_{1b}J_2^3\cos(\theta_x)^3C_{\bf{k}}D_1-4J_{1b}^4D_1^2
+32J_{1b}^2J_2^2\cos(\theta_x)^2\cos(4\theta_y)D_1^2 \\ \qquad
-64J_2^4\cos(\theta_x)^4D_1^2+16J_{1b}J_2\cos(\theta_x)C_{\bf{k}}^3D_2
-4J_{1b}^2C_{\bf{k}}^2D_1D_2-16J_2^2\cos(\theta_x)^2C_{\bf{k}}^2D_1D_2-
16J_{1b}J_2\cos(\theta_x)C_{\bf{k}}D_1^2D_2 \\ \qquad
+4J_{1b}^2D_1^3D_2+16J_2^2\cos(\theta_x)^2D_1^3D_2-C_{\bf{k}}^4D_2^2
+2C_{\bf{k}}^2D_1^2D_2^2-D_1^4D_2^2 ,
\end{array}
\right.
$$
\end{widetext}
where $C_{\bf{k}}$, $D_1$, $D_2$, $\theta_x$, and $\theta_y$ are defined in Eq. \eqref{eq:Fe1.5-H} in the text.

\end{document}